\documentclass[11pt]{article}

\usepackage[utf8]{inputenc}
\usepackage{authblk}      
\usepackage[most]{tcolorbox}
\usepackage{float}
\usepackage{graphicx}
\usepackage{caption}
 \usepackage[sort]{natbib}
\usepackage{pdflscape}

\title{Adaptation by Cumulative Selection}
\author[1]{Rudy Arthur\thanks{Corresponding author: \texttt{R.Arthur@exeter.ac.uk}}}
\affil[1]{Department of Computer Science, University of Exeter, Exeter EX4 4PY, UK}

\date{\today}

\begin{document}

\maketitle

\begin{abstract}
 \noindent Biological systems like long-lived clonal organisms, holobionts and clades challenge traditional evolutionary thinking since they adapt without populations or reproduction. This paper aims to provide an overarching theoretical framework which encompasses standard Darwinian evolution as well as other processes of adaptation. This framework is cumulative selection and I provide a general `recipe' for it to occur. Lewontin's recipe for evolution by natural selection is shown to be a particular example of cumulative selection, but not the only one, with reproduction and populations just one way to perform cumulative selection. I discuss several other examples including clonal organisms, dioecious populations, holobionts, ecosystems, Gaia and neural networks. 
\end{abstract}

\section{Introduction}\label{sec:introduction}

Persistence based selection \citep{bouchard2004evolution, bouchard2011darwinism, ford2014natural, charbonneau2014populations, bourrat2014survivors, toman2017stability, papale2021evolution} has been gaining attention as a way to provide an evolutionary explanation for biological systems like holobionts \citep{doolittle2017s}, clonal organisms \citep{bouchard2008causal}, insect colonies \citep{bouchard2014ecosystem}, Gaia \citep{lenton2018selection} or even taxonomic categories like clades \citep{doolittle2017making}. These examples don't comfortably fit in either the traditional Darwinian paradigm or more expansive frameworks like multilevel selection \citep{okasha2006evolution} and niche construction \citep{odling1996niche}. The problems are
\begin{enumerate}
    \item Identifying a Darwinian population \citep{godfrey2009darwinian}.
    \item Understanding reproduction.
\end{enumerate} 
For example, the well known objection to Gaia, made by Dawkins \citep{dawkins1982extended} and Gould \citep{gould2002structure}, is that there is only one of her (no population) and she doesn't generate `daughter planets' (no reproduction), hence evolution of Gaia, by definition, is impossible. 

Persistence based selection and the related idea \emph{It's the song not the Singer} (ITSNTS) \citep{doolittle2018processes} try to address this by replacing reproduction with the idea that the same `patterns' must persist over time. For standard biological evolution by natural selection (ENS) the pattern is a phenotype and the mechanism of persistence is the copying of genotypes through reproduction. However there are other types of persistence. Individual plants like seagrasses \citep{bricker2018mega} or the quaking aspen grove Pando \citep{pineau2024mosaic} can persist for thousands or tens of thousands of years through clonal propagation. Ecological patterns can persist over geological time, the modern nitrogen cycle has persisted for around 2.7 billion years \citep{canfield2010evolution}, though the species which carry out each function have changed significantly. 

Prior work on persistence based selection addresses the reproduction problem by recognising  reproduction as but one way for patterns to persist. While there are criticisms of this programme \citep{zhang2025defending, papale2021evolution}, it is well now established. Less discussed is the population problem (though see \citep{bourrat2023pricean}). In the case of a clade, or an aspen grove or Gaia, what does it mean to `select' and what is the `population'? Even advocates of a more expansive view of evolution have argued that natural selection is an \emph{``intrinsically population level''} process \citep{sober1984nature}. Survival of the fittest implies multiple individuals and evolution is often \emph{defined} as changes of gene frequencies in populations \citep{dobzhansky1937genetics}.

ENS is an explanation for adaptations in populations of differentially reproducing individuals e.g. long legs allow better evasion of predators allowing long legged variants to leave more offspring. The goal of this paper is to construct similar adaptive explanations for complex biological patterns in systems which don't have standard populations or reproduction. To do this, rather than extending ENS, say by redefining reproduction or population to something other than their usual meanings, I propose to \emph{generalise} it. That means identifying a larger category of processes which can generate and explain complexity, with populations and reproduction one, but not the only, way to perform this general process. This does not mean these adaptive explanations are correct, just as long legs might be due common descent from a long legged founder, but the point is to be able to make adaptive hypotheses in these non-standard situations.

There are many examples of fruitful generalisation in science, especially in mathematics: complex numbers, topological spaces, category theory to name a few. In biology the whole science of taxonomy consists of identifying similarities between entities and grouping them into higher level categories. This paper proposes a (modest) taxonomy of \emph{processes}. ENS, and other process to be discussed, are examples of a general process which I will call \textbf{cumulative selection} \citep{neander1995pruning, neander1995explaining}.

\begin{figure}
    \centering
    \includegraphics[width=0.95\linewidth]{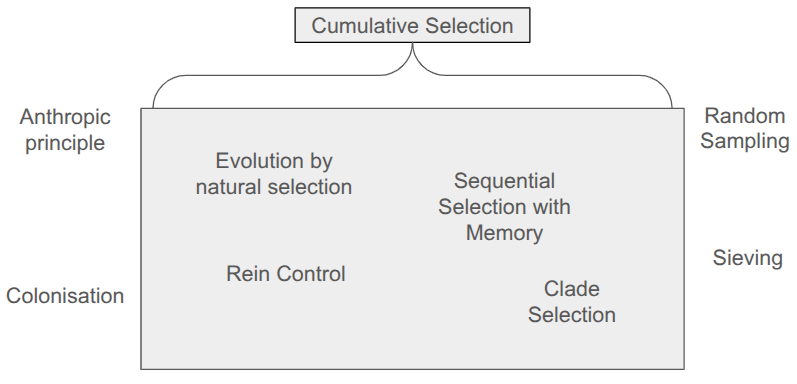}
    \caption{Cumulative selection encompasses numerous processes which can generate complexity. Other processes may well occur in some cases of interest, but are not instances of cumulative selection. }
    \label{fig:diagram1}
\end{figure}
Of course if ENS was the only example of cumulative selection (CS) this would be pointless abstraction. On the other hand, if \emph{any} process of change was an example of cumulative selection, the category would be irrelevant.  I will argue neither of these are the case. There are examples of processes of change which are not cumulative selection and examples of cumulative selection which aren't ENS. I will argue that the latter are just what is needed to describe the development of complexity in systems like holobionts or Gaia. Figure \ref{fig:diagram1} shows a conceptual diagram, highlighting some examples of processes to be discussed at length in the paper.

Section \ref{sec:cumsel} provides conditions for cumulative selection. These are modelled after Lewontin's recipe and reduce to it when CS is applied to populations. In Section \ref{sec:model} I introduce a model first discussed by Dawkins \citep{dawkins1986blind} and Neander \citep{neander1995explaining}. This model is analysed in some detail since it describes a process which is cumulative selection but is not standard ENS. The crucial point of this analysis is that it shows very explicitly that cumulative selection can result in `complex adaptations' (which can be rigorously defined) in a `reasonable' time (sub-exponential). This is what makes cumulative selection a \emph{plausible} explanation of observed complexity. Readers in a rush to see other examples of CS can skip Sections \ref{sec:complexity} and \ref{sec:randomsampling}, although I believe these provide useful insight into the explanatory power of cumulative selection. In Section \ref{sec:examples} I discuss numerous other examples of change which are and are not examples of cumulative selection. These are listed in Tables \ref{tab:neg} and \ref{tab:pos}. I conclude in Section \ref{sec:disc} by summarising the advantages of this perspective.

\section{Cumulative Selection}\label{sec:cumsel}

Suppose there is some \emph{entity} or \emph{system} that has one or more \emph{features}. The particular \emph{values} of the features at some instant are the system's \emph{state}. If
\begin{enumerate}
    \item Feature values vary randomly over time (variation)
    \item Different states have different rates of persistence (differential persistence)
    \item The persistence of the current state is similar to the persistence of recent states (memory)
\end{enumerate}
then the system is subject to cumulative selection on its features. There has to be variation otherwise there is nothing to select (1). There have to be different rates of persistence, otherwise there is nothing to select for (2). There cannot be too much variation, otherwise the process cannot be cumulative (3). Particular applications will give concrete meanings to the terms `system', `feature' and `value' as well as identifying the source of variation, the factors that determine the persistence of a state and the precise meaning of `recent'. 

The persistence of a state or feature refers to how long it takes to change its value(s) by some set amount. I define an `adaptation' as a feature value, or set of feature values, which increase the persistence of a state. The basic loop of cumulative selection is
\begin{itemize}
    \item A feature that varies to lower its persistence will vary again soon.
    \item A feature that varies to raise its persistence will take longer to vary again.
\end{itemize}
Condition 3 implies that the persistence of each variation is not random, but similar to the persistence of the prior state. We assume persistence can increase or decrease with each variation. A persistence decreasing trajectory experiences more and more variation on an accelerating timescale until it hits a persistence increasing variation. This biases the process against persistence decrease. This bias together with the observation that, \emph{ipso facto}, the system spends a long time in persistent states, means that when observing the state of the system we will tend to find it in a persistent state. Thus cumulative selection for persistence could explain why we observe a system in a given state.

On a persistence increasing trajectory as states grow more persistent it takes longer and longer to get to the next one. We could therefore reasonably ask if CS is more efficient at generating persistent states than randomly guessing feature values. The model developed in Section \ref{sec:model} will show that it can be \emph{much} more efficient, so CS is not only a possible explanation but a \emph{plausible} one.

\subsection{Connection to ENS}

Lewontin's recipe, quoting directly from \citep{lewontin1970units}, is
\begin{enumerate}
\item Different individuals in a population have different morphologies, physiologies,
and behaviours (phenotypic variation)
\item Different phenotypes have different rates of survival and reproduction in
different environments (differential fitness)
\item There is a correlation between parents and offspring in the contribution of each to
future generations (fitness is heritable)
\end{enumerate}
\begin{quote}
 These three principles embody the principle of evolution by natural selection. While they hold, a population will undergo evolutionary change. 
\end{quote}
 Although the generality of these conditions is challenged, mostly over the inclusion or exclusion of genetic drift \citep{brandon2014adaptation}, and alternatives exist \citep{hull1980individuality}, this recipe is generally accepted as a reasonable précis of ENS \citep{okasha2006evolution, papale2021evolution, zhang2025defending}. In this section I will show how the conditions for cumulative selection applied to reproducing populations recover ENS as described by Lewontin's recipe.

The thing-to-be-explained is the prevalence of a phenotype among a group of individuals, say, the beak shape of some Galápagos finches. The `system' here is a Darwinian population, a collection of individuals which perform interbreeding and reproductive competition \citep{godfrey2009darwinian}. The `features' are the particular phenotypes of the individual members of that population and the `state' is a list of the phenotypes of its members. If those phenotypes are quantitative (beak size) one could literally write down the state as a list of numbers or summarize it through population level statistics like average beak size. CS Condition 1 is met by phenotypic variation, by the definition we just gave of a feature as a phenotype and so is equivalent to Lewontin's first condition. 

The way that persistence is realised is through survival and reproduction of individuals in the population, with, say, the long beaked variety more likely to  be `copied' into the future state. To be very explicit, if the only feature of interest is beak size and there are two varieties, small and large then the state of the population is completely described by one feature, $p(t)$, the frequency of the long beak variety at time $t$. If the fitness advantage of a long beak is $s$ then standard arguments give, for discrete generations,
$$
\Delta p =\frac{sp(1-p)}{1+sp}.
$$
For weak selection we can make a continuous approximation which is easier to analyse,
\begin{equation}\label{eqn:fitnesslogistic}
    \frac{dp}{dt} = sp(1-p).
\end{equation}
We define persistence as the time a state remains within some fixed neighbourhood $\epsilon$ of its initial value, i.e. the time required to go from $p \rightarrow p+\epsilon$. This is
$$
T_\epsilon = \int_p^{p+\epsilon} \frac{dx}{sx(1-x)} = \frac{1}{s} \left( \ln \frac{p+\epsilon}{p} - \ln \frac{1-p-\epsilon}{1-p} \right)
$$
For small $\epsilon$
$$
T_\epsilon \simeq \frac{1}{sp(1-p)}
$$
i.e. persistence is inverse to the rate of evolutionary change. The important fact is that, persistence is function of fitness, $s$, so one determines the other. This is all we need to demonstrate that CS Condition 2, differential persistence, is equivalent to Lewontin's second condition, differential fitness.

CS Condition 3 covers the rate of variation. While variation is required for change, it cannot be unbounded. This is exactly what \cite{godfrey2009darwinian} refers to as the `error catastrophe'. If long beaks are more likely to survive and reproduce, but the long beak trait is not reliably passed to the offspring, then the state will just `bounce around' rather than converge on a high fraction of long beaks. Feature/phenotype variation is required for change, but too much and it cannot be \emph{cumulative} change. In this case, CS condition 3 with `recent'  understood as `in the previous generation' is equivalent to requiring correlation between parents and offspring, and so is equivalent to Lewontin's third condition.

Recovering Lewontin's recipe implies evolution by natural selection is an example of cumulative selection and achieves two important things. First, it is a sanity check, ENS is an example of cumulative selection as claimed. Second, ENS also displays decelerating convergence. Equation \ref{eqn:fitnesslogistic} is the logistic equation, so fixation of a beneficial allele follows an $s$-curve, showing that the deceleration is not a barrier to adaptation. 

This is not an argument that CS should be preferred to ENS or is an alternative to it. By specifying more information about the system, in this case that it is a population of reproducing individuals, ENS provides more detail than CS. The explanatory goal determines if this extra information is needed. To make an analogy, one might explain some trait of an organism as due to it being a eukaryote, a plant or a fern. As we get more specific, we can explain more detail about the organism. Similarly, identifying a process as cumulative selection tells us if that process can plausibly generate adaptations through random variation. Specifying more detail about that process (how it is carried out and by who) gives us a more detailed theory.

\section{A model of selection}\label{sec:model}

In this section we study, in some detail, a model which is simple enough to analyse completely which demonstrates adaptation without populations or reproduction. It is based on a model described in \citep{dawkins1986blind} and \citep{neander1995pruning}. For Dawkins' version, consider a device like a combination lock with 23 wheels which can each be spun to display the letters A-Z. Starting from a random sequence the goal is to end up on the target sequence 
$$METHINKSITISLIKEAWEASEL$$ 
Dawkins implements a kind of sequence `breeding' but we are just going to spin the wheels randomly. \citep{neander1995pruning} describes the same model using the example of a lottery. A random number sequence is drawn, consisting of 7 numbers between 1 and 30 and she wants to find a matching ticket. Both Dawkins and Neander contrast the strategy of randomly picking sequences, which has an extremely low probability of finding a match, against a \emph{cumulative} strategy. In the cumulative strategy whenever a correct letter or number is found, it is fixed and this strategy has a much higher chance of finding the target.

`Extremely low' and `much higher' do not do justice to the difference between the two situations. If $p$ is the probability of a match at one position and $L$ is the length of the sequence, the expected waiting time for a random search is $1/p^L$. For a cumulative search the \emph{worst case} is $L/p$ (see Appendix \ref{sec:app}). So, for Dawkins' combination lock, if we randomly guessed one combination per second we would expect to have to try for $10^{25}$ years to get the target phrase, equal to around one quadrillion times the age of the universe. With the cumulative strategy, we would expect to guess it in about $10$ minutes. 

Generalising Dawkins and Neander's examples, consider $N$ sequences of length $L$, made using the characters from an alphabet of size $b$. Fix a sequence, called the target, which we aim to match. Note, neither, I, Dawkins or Neander are claiming ENS searches for a specific target, teleology. Rather this models the inverse problem, how long would it take to evolve something complex and unlikely (an eye) from humble beginnings (a light sensitive membrane)? This is essentially a model of search on a  fitness landscape \citep{gavrilets1997evolution} used as a tractable example that has some similarity to processes of biological change.

We approach the target using a cumulative or \emph{ratcheting} search (for discussion of other evolutionary ratchets see \cite{lukevs2011neutral, tennie2009ratcheting, arthur2022selection}). This means that at some moment we have a `state' or template consisting of $k$ matches and $L-k$ mismatches, for example,
$$MExxxxNKSxTISLxxEAxxASxL$$ 
We generate $N$ new sequences by copying the current best state. We change the matched characters with probability $\nu$ and change the mis-matched ones with probability $\mu$. We then choose the sequence with the most matches as the template for the next round and repeat the process until the target is reached. In the case of a tie, two sequences with different matches but the same number, we pick one at random to be the template. Dawkins' model would correspond to $L=23$, $b=26$, Neander's to $L=7$, $b=30$ and both to $\nu=0$, $\mu=1$. A random search would have $\nu = \mu = 1$.

We will mostly consider the case $N=1$, where we don't have a population so the best state is always the current one. No reproduction is required, in the combination lock version we just spin all the wheels every time-step, there is no need to destroy the old lock and make a new one. At first we will examine the case with no `back-mutation', $\nu = 0$. Consideration of $\nu \neq 0$ will be important for defining the equivalent of heredity. I will refer to this model as the lottery model after \citep{neander1995pruning}.

\subsection{Ordered versus unordered paths}

In the famous `Spandrels of san Marco' paper, \cite{gould1979spandrels} argued that certain observations might be better described as being due to developmental constraints rather than ENS. There is a problem with a similar flavour in this simple model, where developmental constraints might slow or even prevent us from reaching the target. Consider a transformation taking the word $HEAD$ to the word $TAIL$, one letter at a time. If we allow only valid words at each step a possible path is
\begin{align*}
&HEAD\\
&HEAL\\
&TEAL\\
&TELL\\
&TALL\\
&TAIL\\
\end{align*}
This is an example of a word ladder\footnote{Invented by Lewis Carroll \citep{carroll2016doublets}}. A much simpler task, which is the version considered by Dawkins and Neander, permits any string at the intermediate stages
\begin{align*}
&HEAD\\
&TEAD\\
&TAAD\\
&TAID\\
&TAIL\\
\end{align*}
I will refer to these as the `ordered' and `unordered' cases respectively. An ENS interpretation would replace `valid word' with `viable offspring'.  A path always exists in the unordered case but a path might not exist or be extremely long in the ordered case and paths in real fitness spaces are quite complex \citep{gavrilets1997evolution}. We will examine both the ordered and unordered versions and for the ordered version consider the worst feasible case, where there only one viable path.

\subsection{Model solutions}\label{sec:solutions}

We solve the random and cumulative versions of the model for $N=1$, $\nu = 0$ in Appendix \ref{sec:app}. The basic result is the \emph{time to discovery} $T$, the expected time it takes to land on the target sequence which has $L$ characters different from the starting one.
\begin{equation}
   T_{random}(L) =  b^L 
\end{equation}
\begin{equation}
   T_{ordered}(L) =  \frac{L}{p} - (L-1)
\end{equation}
\begin{equation}
   T_{unordered}(L) \simeq \frac{\ln L + \gamma}{p}
\end{equation}
Where $\gamma$ is a constant and $p = \frac{\mu}{b}$. For $N>1$ the solution for the random search is
\begin{equation}
   T_{random}(L, N) =  \frac{b^L}{N}
\end{equation}
The analytical solution of the cumulative search for $N>1$ is harder, but computer simulation is simple and some plots are shown in Appendix \ref{sec:app}. 

The key observations are
\begin{enumerate}
    \item The ordered search is exponentially faster than the random one.
    \item The unordered search is exponentially faster than the ordered one.
\end{enumerate}
The ordered and unordered cases represent upper and lower bounds on the cumulative search time. In most situations we would expect more than one viable path e.g. using synonymous codons, but  not for every path to be viable. However, even in the worst case, the ordered search only takes time linear in $L$ compared to the exponential time for the random search. The upshot is that cumulative selection can, in sub-exponential time, discover complex adaptations through random variation, \emph{even with a population of one}. 

\subsection{ENS or CS?}
Consider Lewontin's recipe. Lewontin's first condition refers to `different individuals in a population'. While we can use many tickets, we can still reach the target using only one, as long as we can modify that ticket. Lewontin's second condition refers to survival and reproduction. In the lottery model, it is irrelevant if, for each trial, the numbers on the current lottery ticket are copied to a new one, asexual reproduction, or the same ticket is used with the incorrect numbers erased and replaced, modification. Lewontin's third condition refers to heredity and again explicitly references reproduction. A single ticket subject to random variation does not find the target sequence by Lewontonian ENS.

On the other hand it does satisfy the conditions for cumulative selection. The system is the ticket, the features are the letters in the sequence and the state is some particular list of letters. The features vary, matched characters at a rate $\nu$ and unmatched at a rate $\mu$. For the unordered case, if there are $L$ sites left to match, the chances of at least one match in the next step is $1 - (1-p)^L$, a monotonically increasing function of $L$. So a state with 4 correct letters and $L-4$ remaining will persist longer than a state with 3 correct and $L-3$ remaining, satisfying the second CS condition (the ordered case is tricker, see Appendix \ref{sec:app}). 

For the third condition, recall we allow back-mutation at a rate $\nu$. For a site with a correct character the probability for it to remain correct in the next generation is
$$
r = (1-\nu) + \frac{\nu}{b}
$$
meaning either it doesn't mutate, or it `mutates' but to the correct character, for example a synonymous mutation that takes the character $W$ to the character $W$. For very small $\nu$ everything remains the same, though now matched characters are very long lived, rather than immortal. However, if the probability of back-mutation is high then there is no persistence of state over time. There needs to be enough continuity for the process to actually be cumulative, we can't build on previous matches if those matches are constantly being lost. In the lottery model we define the \emph{error threshold} as the critical value $\nu_c$ above which there is no cumulative selection and the search becomes effectively random.  This threshold is computed for the ordered and unordered models in Appendix \ref{sec:app}. Interestingly, in the ordered case it is independent of $\mu$. As long as we remain below the error threshold, CS Condition 3 is satisfied and cumulative selection occurs even though ENS does not.

To address some possible objections, first, it might be argued that this is really ENS, if we just reinterpret things e.g. modification is a type of reproduction \citep{papale2021evolution, doolittle2024darwinizing} or the possible feature values are a kind of abstract population. If the question is ``how could this sequence arise by random variation?'', this is unnecessary. We can take the plain meaning, there is one ticket that gets modified. This system satisfies the conditions for CS, which is all that is needed for an adaptive explanation. Second, it might be argued that this model is teleological because the correct string is known in advance. Reducing $N$ to $1$ in Dawkins' and Neander's model does not introduce teleology. It is optimisation on a fitness landscape with a single peak. In realistic scenarios the fitness landscape is more complex and determined by environmental conditions and interactions.

Lastly, \citep{sober1984nature} and others seem to, definitionally, require a population for selection. The process described here doesn't need a population and yet the winning ticket or correct phrase is `selected' by it. Following \citep{neander1995explaining} and others \citep{nanay2005can} who use the term `cumulative selection' in relation to this model, calling this process selection is apt. The word selection only implies multiple options rather than a population e.g. `selecting' an item from a menu or a frequency on a radio can be done with one copy of the menu and one radio! When applied to populations, CS is ENS and selection has its usual meaning in that context. In other cases it has an analogous interpretation as a way of biasing a random process towards certain outcomes, namely, persistent ones.

\subsection{Complexity}\label{sec:complexity} 

Having a recipe for cumulative selection and a concrete model allows us to be a bit more precise about what we mean by a `complex' adaptation. There are many definitions of complexity \citep{adami2002complexity}, usually relating to the diversity or the number of parts of an entity and often quantified using information theory and compression \citep{nordin1995complexity}. I will call this `structural complexity'. With any compression based measure of structural complexity the string 
$$
AAAAAAAAAAAAAAAAAAAAAAA
$$
is less complex than the string
$$
METHINKSITISLIKEAWEASEL.
$$

In the sequence model however, 
$$METHINKSITISLIKEAWEASEL,$$ 
$$AAAAAAAAAAAAAAAAAAAAAAA,$$ 
and any other 23 character sequences are just as hard to find from a random starting point. I will call an adaptation `complex' when a random search would take an extremely long time to discover it. Obviously `extremly long' depends on the system under consideration, but usually this can be taken to mean much longer that the lifetime of the system. To distinguish this notion of complexity from structural complexity I call it `search complexity'. In the lottery model a random search takes time $O(b^L)$, unreasonably long for large $L$. 

For adaptation of a continuous parameter we can think of binning that parameter so that all values in a bin $[s_i, s_i +\delta s_i]$ are equivalently persistent, e.g. states in a bin persist to the same extent within some small range $\pm \epsilon$.  If a large range of parameters is equivalent in persistence we don't need many bins and so can find the best one quickly by random search. If only a very small range of parameters is maximally persistent there are many bins and it becomes very hard to find the right one at random. I will call such an adaptation `fine-tuned'. 

The search complexity depends on where we start. If we have already matched 22 out of 23 characters, the final adaptation is not search complex, if we start from 0 it is, even though both targets can be equally \emph{structurally} complex (indeed they can be the same). Likewise, a continuous parameter that only needs to change slightly is not fine-tuned. I consider this initial condition dependence a desirable feature of the definition. It means a slight increase in limb length is not a complex adaptation, but evolving two limbs from no limbs certainly is! The magnitude of change and the timescale for making it matter when we are asking if an observation can be the result of a random process. Cumulative selection is a way that search complex or fine-tuned adaptations can be discovered in reasonably short times via random variation. 

We earlier discussed slowing down as adaptation occurs. This also occurs in the unordered lottery model. If there are $L$ sites left to match, the chance of at least one match is $1 - (1-p)^L$, which decreases with decreasing $L$. For much more discussion on this slowing down as the target is approached, see \citep{arthur2022selection} and \citep{arthur2023does}. Making these arguments rigorous for continuous parameters would take us into the realm of stochastic calculus \citep{ewens2004mathematical}, which I leave for future work. Here I simply note again this slowing down does not prevent cumulative selection from discovering the target much faster than a random search.

Finally, I note that some authors \citep{mcshea2010biology, wong2023roles}, have proposed growth in \emph{structural} complexity as a fundamental law of nature. Despite arguing elsewhere that increasing structural complexity is important for Gaia Theory \citep{arthur2022selection, arthur2023does}, increasing structural complexity is not an inevitable outcome of cumulative selection. If structurally complex states are increasingly persistent, structural complexity will be selected. This is what I have argued elsewhere holds for Gaia. Standard ENS provides examples where structural complexity is selected against. One example is genome-streamlining \citep{hessen2010genome}, which reduces the length of DNA sequences. There are many other well known cases where organisms lose function, and therefore complexity, but still gain fitness. For example the loss of sight in cave dwelling fish \citep{moran2015energetic}. There is no contradiction with cumulative selection, if shorter genomes persist better than longer ones they will be selected. Cumulative selection will discover structurally complex states only insofar as structural complexity increases persistence. 

\subsection{ Random Sampling }\label{sec:randomsampling}

Finally, let us cover the non-cumulative version of the lottery model in more detail. Let $N=1$ and take Dawkins' example where $b = 26$. If we can afford one billion random trials we can expect to discover a sequence of length
$$
L = \frac{\log T}{\log b} \simeq 6.
$$ 
The exponential growth of the discovery time means that finding \emph{complex} adaptations (here target sequences with large $L$) by random search is hopeless. An explanation of an adaptation that relies on its \emph{de novo} emergence, a `hopeful monster', is improbable. 

This cannot be evaded by having a large population. The size of the population is a multiplicative factor to the number of steps
$$
L = \frac{\log ( TN )}{\log b} 
$$
So, with a billion individuals doing a billion independent searches and $b=26$ we can expect to find a sequence of length $L=12$. Having many individuals is not enough to compensate for the huge times required to discover sequences of even moderate length. This also shows that small mutation rates in the cumulative models, where in the worst case $T = Lb/\mu$, are not enough to give random searching an advantage for long sequences (see also Figures 1 and 2 in the Appendix).

This point is why anthropic explanations \citep{smolin2007scientific} are bad ones. Although these arguments are perhaps well-known in evolutionary biology, there are other fields where they not widely appreciated, particularly the study of planetary regulation and astrobiology. For example, the idea of `lucky Gaia' \citep{watson2004gaia} is unviable. Put simply this is the proposal that the universe is quite old and there are a lot of planets, so one would be expect there to be a habitable one by chance alone. Since there are roughly $10^{12}$ stars in the Milky Way and evidence for at least one planet around each \citep{cassan2012one} this large number would seem to make anything possible. 

However, as shown above, the complexity of what can be found through random, non-cumulative searching depends on the \emph{log}
of the population and $\log 10^{12} = 12 \log 10$ is not so big. Gaia, a planetary regulation system, consisting of interconnected biogeochemical cycles is surely quite complex, in both search and structure. Even astronomically large numbers are not enough to `beat the log' and simply having a very large number of planets is not a guarantee of winning the Gaian lottery. Rather than luck, we need mechanisms. It may be that there are many Gaian states, or Gaian states are `attractors', or cumulative selection is operating, see \cite{arthur2022selection} for more discussion on this issue. In general, complex observations demand explanations that don't depend on luck.

\section{Examples of Cumulative Selection}\label{sec:examples}

\subsection{Negative examples}

We showed in Section \ref{sec:cumsel} that cumulative selection encompasses Lewontin's recipe and thus evolution by natural selection, insofar as Lewontin's recipe does. More interesting are systems which are not Darwinian populations but do satisfy the cumulative selection conditions. I will give examples of such systems in the next Section but it is also useful to look at cases where the recipe does not hold.
\begin{enumerate}
    \item If the system is an (idealised) individual organism, a fruit fly say, with genes as features then it passes conditions 2 and 3 but the genome is fixed, failing condition 1. 
    \item If the system is a population with individuals that vary in a way which provides no difference in persistence then this passes conditions 1 and 3 but fails 2.
    \item A random search passes conditions 1 and 2 but fails condition 3.
\end{enumerate}
Regarding the first example, implicit in our, and Lewontin's, conditions is that the system in question, population or individual, does not go extinct, which \cite{papale2021evolution} refers to as regeneration. While environmental radiation could change the genes of an individual, this will likely kill it. We could add as a $0^{\text{th}}$ condition for cumulative selection that variations don't kill the system, but allowing the possibility that variations can end the game is sometimes important, see \citep{arthur2023gaian} for a concrete example.

Regarding the second example, I note \cite{brandon2014adaptation} claims that genetic drift satisfies Lewontin's conditions for ENS and so would also be an example of CS. \cite{brandon2014adaptation}'s point is that in a finite population, by chance, some neutral mutation will predominate, by Lewontin's Condition 3 this predominance will be passed to the next generation, resulting in a random walk that will eventually lead to fixation of a neutral trait. 

From the point of view of cumulative selection there are two answers. We could accept the argument, allow genetic drift as a type of ENS and therefore as a type of CS. A more satisfying answer begins by noting \cite{brandon2014adaptation}'s differential fitness condition is \emph{Different variants leave different numbers of offspring in succeeding generations}. The more specific version I use in Section \ref{sec:cumsel}, from \citep{lewontin1970units}, is \emph{Different phenotypes have different rates of survival and reproduction in different environments}. By definition, in a neutral environment different phenotypes have the same \textbf{probability} of survival and reproduction. This is true regardless of what actually happens in a finite population. The second conditions of both Lewontin's recipe and cumulative selection should be understood as specifying that the reproduction probability/persistence is not a constant function of the phenotype/feature value, definitionally excluding drift. This is just the \emph{propensity interpretation} \citep{popper1959propensity} which is even advocated by \cite{brandon2014adaptation} himself.

The final important negative example to discuss is the progressive sieve \citep{neander1995pruning}. A progressive sieve is a cylinder divided along its length by a series of mesh grids that get finer and finer. Imagine pouring gravel or different sized beads in at the top. As the beads pass from top to bottom they would appear to be cumulatively selected, until only the smallest ones make it to the bottom. If the entity is taken to be an individual bead, there is no variation hence no cumulative selection (failing condition 1). If the entity is taken to be the whole collection of beads then there is no \emph{random} variation also failing condition 1. Van Valen's oft quoted example of granite weathering which preserves quartz crystals but dissolves feldspar and mica to create beaches \citep{van1989three} is, quite literally, a progressive sieve and has also been called weak ENS in the biological context \citep{okasha2006evolution, bourrat2014survivors}. \emph{Cumulative} selection applies to systems where variation is continually generated and is not simply a process which filters a fixed population.

With this in mind, consider Selection by Survival, the simplest formulation of which is \emph{``Natural selection in a contracting population of non-reproducing, non-competing, individuals''} \citep{ford2014natural}. Stated as such, this is an example of a progressive sieve, and not cumulative selection. However, the surviving individuals (Doolittle has something like biological clades in mind) are often assumed to change over their lifetime. If the subject of cumulative selection is taken to be a clade and one of its features its diversity then 1. diversity varies randomly over (long enough) time through speciation 2. high diversity clades (likely) persist longer than low diversity ones and 3. a clade is, by definition, still the same clade regardless of the rate of variation. Thus clade diversity \emph{can} be subject to cumulative selection and cumulative selection could explain observations of high clade diversity. 

Finally, note, the fact that a progressive sieve is not an example of cumulative selection does not mean such sieves don't operate or have no observable effect. See \citep{arthur2023gaian} which explicitly studies a system with simultaneous sieving and cumulative selection.

\subsection{Positive examples}\label{sec:positiveexamples}

\subsubsection{Pando} Consider a large, long-lived, clonal organism such as the quaking aspen Pando. This is our entity. We map the discussion of \cite{bouchard2008causal} to our conditions. Let the features of Pando correspond to its \emph{ramets}, in particular, their location. The state is the location of all the individual ramets, which we could summarise through the `shape' of the grove and the density of trees. The `target' or explicandum is the observed shape and density of Pando.

The ramets vary in their location, with new ramets appearing at new locations over time through suckering from the main root system and old ramets dying (condition 1). Ramets at different locations are more or less persistent, due to soil conditions, vulnerability to grazing and so on (condition 2). A `daughter' can only grow 30-40m from a `mother' ramet, which is small compared to the total size of Pando at about 40 hectares (condition 3). As the three conditions are satisfied, the shape and density of Pando can be subject to cumulative selection. Note that virtually identical arguments could be applied to any asexually reproducing species, as discussed by \citep{bouchard2004evolution}, since all members are, similarly to ramets, clones of their common ancestor.

There are several points which complicate this example. First I note that if ramets could grow arbitrarily far away from their parents, we would still have a changing system, with ramets persisting longer depending on where they ended up, however it would not be cumulative selection. An area of favourable soil might support higher density, but if the daughter ramets just popped up anywhere, they would be relying on luck to land in that patch and there would be no consistent increase in the density there over time as old trees die and their `offspring' are not nearby to replace them. The exact transition between cumulative selection and random search depends on the range of daughter ramets is an interesting question for future work.

Second, clonal organisms, like Pando, also experience somatic mutation which may have consequences for the ramets \citep{clarke2012plant}. This would be a different type of variation than ramet location and should be analysed separately. Indeed aspens are also dioecious, and this type of reproduction fits well within the standard theory of ENS \citep{tuomi1989hierarchical}. The discussion above, following \cite{bouchard2004evolution}, refers to an idealised case where ramets are perfect clones and production of new colonies or competition between neighbouring colonies is not relevant.

Lastly, and most crucially \citep{papale2021evolution}, following \citep{godfrey2009darwinian} argues that even the idealised case descried by \citep{bouchard2008causal} does not really require a new formulation of ENS. Lewontin's recipe doesn't mandate new genetic individuals and certainly applies to asexually reproducing bacteria, so why not to ramets? I would agree with \citep{papale2021evolution} that Pando alone does not compel one to abandon standard ENS.

This is not a technical problem for cumulative selection, ramet evolution is a fine example as shown. However if this can also be explained as ENS, what is added by cumulative selection? I provide two answers. First, even if Pando's `shape' can be explained by ENS applied to ramets this is still an unusual application of the theory of ENS and, despite \citep{papale2021evolution}'s claims, applying ENS to plants, or even counting individuals, is a very difficult and much debated question, see \citep{clarke2012plant}. On the other hand, to provide an adaptive account of observations it is enough to know only that the ramets are cumulatively selected. Second, if Pando was the only example I would agree there is no need to consider a more general process. However there are other examples we have already seen (cumulative search in the lottery, clade selection) and more that we will see in this section that are not reducible to standard ENS.

\subsubsection{Sex Ratio} The standard game theoretic argument for 50:50 sex ratios in dioecious species is \citep{hamilton1967extraordinary}: If male births are less common, a newborn male has better mating prospects than a newborn female, therefore parents that have more male offspring have more grandchildren, spreading the genes for male production, reducing the male advantage. The same holds for a deficit of females. Therefore the system is only stable with an equal ratio of males and females. 

This is an individual level argument which is correct but we can describe the same situation in a different way. Let our system be a population and one of its features its sex ratio. The state is the value of the sex ratio. Sex ratio varies randomly due to inherent variance in the numbers of births and deaths (condition 1). States close to 50:50 persist for longer than ones far from 50:50 (condition 2). Variations are relatively small i.e. we do not leap from a 10:90 ratio to 90:10 in one generation (condition 3). This means we could consider sex ratios subject to cumulative selection acting on a population. 

The individual argument is perfectly valid, but if we want to explain an intrinsically population level phenomenon, sex ratio, with a population level selection argument, in accord with Williams's principle \citep{sober2011adaptation}, this is a way to do so. Williams's principle is perhaps less compelling for the sex ratio, which is simply an aggregate of individual character values \citep{grantham1995hierarchical}. This application of CS is mainly illustrative, showing that the fine tuned value of a feature, sex ratio, can be accounted for by CS. It does not provide any more insight than Hamilton's account, which uses more detailed information about the system. I now turn to cases which I believe do show advantages of the CS perspective.

\subsubsection{Holobionts}
A microbiome refers to the community of micro-organisms within a host and a \emph{holobiont} is the host-microbiome considered as a single entity \citep{margulis1991symbiosis}. Understanding the evolution of holobionts has driven a lot of the recent interest in expanded and alternative formulations of ENS \citep{doolittle2017s, doolittle2018processes, roughgarden2018holobionts, van2019role, papale2021evolution}. Proponents of holobiont evolution claim host and microbiome are sufficiently integrated to create a \emph{hologenome} \citep{zilber2008role} upon which selection can act. Others claim that the host-microbiome pair is better treated as a community or an example of co-evolution, rather than a composite organism and `unit of selection' in its own right \citep{moran2015hologenome, skillings2016holobionts, lloyd2019criteria}. 

I do not propose to resolve this debate, rather I propose that holobiont adaptation may arise via cumulative selection. Let the system be a population of holobionts with features the hologenomes of its members. To be precise I will follow the model of \citep{roughgarden2020holobiont} where the number of microbes within each host is the main variable of interest, so in this case the average microbiome size is the focal feature.

Microbiomes are passed both vertically (from parent to offspring) and horizontally (obtained by the host from the environment). \cite{roughgarden2020holobiont}'s model of vertical transmission consists of repeating the following steps,
\begin{enumerate}
    \item Proliferation of microbes within hosts.
    \item Reproduction based on microbiome size, passing microbiome from parents to offspring.
    \item Mixing step: each microbe has a probability $m$ of exiting its host and entering a global `transfer pool' which is redistributed uniformly at random.
\end{enumerate}
In the model of horizontal transmission, the steps are
\begin{enumerate}
    \item Proliferation of microbes within hosts.
    \item Reproduction based on microbiome size.
    \item Microbes are all sent to a source pool, offspring are sent to a host pool.
    \item Microbiomes are sampled randomly by hosts from the source pool.
\end{enumerate}
The model treats all microbes as having the same effect on the host, parasitic or mutualistic. Hence the number of microbes determines the fitness of the host and its reproductive success. For simplicity, let us choose the version where microbes are mutualistic and therefore hosts with many microbes have higher reproduction rates.  The system is the population, its relevant feature is the average microbiome size and the state is the value at some instant.

In the vertical model CS Condition 1 is met  by the mixing step CS Condition 2 for differential persistence is realised by the hosts' reproductive success being a function of its microbiome size. Condition 3 is met by having a sufficiently small, but non-zero, value of $m$. In the horizontal model CS Condition 1 is met by the random sampling from the source pool. CS Condition 2 is again met because the reproductive success is a function of microbiome size.

So far, we have more or less standard ENS since the CS and Lewontin conditions are the same. However in the horizontal model hosts contribute all their microbiome to the source pool and their offspring randomly sample some number of microbes from it. Thus there is no correlation between parent and offspring in the trait, violating Lewontin's heritability condition 3. However, the average microbiome size is similar to the previous generation, satisfying CS Condition 3 for state similarity.

Thus we agree with e.g. \citep{zhang2025defending} that holobiont reproduction with horizontal transmission is not ENS. Nevertheless it is still CS and hence holobiont adaptation can occur through random variation, which is the key point.

\subsubsection{DaisyWorld}  Let the entity be Gaia, the coupled system of life and environment on Earth. Let us use the famous DaisyWorld model of Gaia \citep{watson1983biological}. In this model the planetary albedo determines how much sunlight is reflected or absorbed, changing the planetary temperature. On DaisyWorld a white/black daisy reflects/absorbs sunlight cooling/heating the local environment and the planet by changing the albedo. A more realistic model might consider cloud or forest cover instead of daisies. 

If the background temperature is hotter than ideal, the growth of black daisies is self-limited, they make it hotter so more black daisies makes it harder for any daisies to grow. White daisies on the other hand, cool, encouraging the growth of daisies. DaisyWorld tends to a ratio of black and white daisies which results in global temperatures close to the optimum for both species. Assuming a high background temperature on DaisyWorld
\begin{enumerate}
    \item Albedo varies (due to random fluctuations in the birth/death of daisies)
    \item States with low albedo do not persist as long as states with high albedo
    \item There are no dramatic fluctuations in the albedo
\end{enumerate}
So albedo is a cumulatively selected feature of DaisyWorld. This process is usually known as \emph{rein control} \citep{harvey2004homeostasis} since white and black daisies are `pulling' temperature in opposite directions. Removing white daisies gives \emph{single} rein control. Black daisies raise the temperature above their optimal growth temperature, until it is high enough that their own reproduction is limited, with the temperature stabilizing at some value between the temperature for optimal growth and the upper limit for reproduction, corresponding to some equilibrium value of albedo $A_{eq}$. The second condition would then change to
\begin{enumerate}
  \setcounter{enumi}{1}
\item States with albedo far from $A_{eq}$ do not persist as long as states close to it
\end{enumerate}
Single rein control still finds states of high persistence, which are not equivalent to states optimising reproduction or biomass. However, either double or single rein control are types of cumulative selection.

In Daisyworld the cause of albedo variation is known, it's the proportion of black and white daisies. Notice the similarity to the sex ratio. In both cases there is a micro-scale description in terms of individual males/females or black/white daisies and a macro-scale description in terms of cumulative selection acting on a feature of the macro-scale entity, the sex-ratio/albedo of a population/Gaia. Also similar to the case of the sex ratio, in Daisyworld the albedo is an aggregate feature and the CS perspective doesn't add a huge amount to the analysis of DaisyWorld.

However, in reality albedo is a much more complicated function of abiotic and biotic interactions which is not reducible to a sum over individual contributions. Nonetheless one can still claim that the albedo of Earth is cumulatively selected. Namely if 1. albedo varies 2. some albedos are more persistent and 3. albedo does not change dramatically on short timescales then the `fine-tuning' of albedo can be explained by CS. By analogy with Daisyworld, the famous CLAW hypothesis \citep{charlson1987oceanic} proposed that that life causes variations in albedo through creation of cloud condensation nuclei. Then if life creates albedo conditions favourable for life, life can continue to maintain the albedo at some favourable value. If life creates unfavourable albedo, it is self limiting. This is exactly the loop described in Section \ref{sec:cumsel}. The CLAW hypothesis provides `micro' level details of how variation, persistence and state similarity are realised for Gaia.

In standard evolutionary theory a consistent adaptationist story does not guarantee its validity, nor does ENS provide the genetic and environmental details resulting in the evolution of any particular trait. It does however provide an extremely useful overarching theoretical framework. Thus, CS alone does not provide the microscopic details of the process. Likewise, \cite{charlson1987oceanic}'s consistent story does not guarantee albedo has been cumulatively selected.  Rather, like ENS, CS is an overarching theoretical framework.

\subsubsection{Ecosystem Evolution} 
Evolutionary ecology usually studies co-evolution of species with each other and their environments \citep{odling1996niche}. In contrast, the theory of ecological fitting \citep{janzen1985ecological}, is a process whereby species colonise new environments or form new interactions using traits they carry at the time of arrival. It can be considered a null model for co-evolution, the `fit' of species to a habitat is not due to species' evolution in that environment, nor to their historic shaping of that environment, but rather their reuse of prior adaptations and exaptations. 

Consider an uninhabited island where a large number of species suddenly arrive. Plants may colonise bare ground, insects can eat those plants, mice can eat those insects etc. I will call this \emph{ecological colonisation}. Though probably not common, at least one artificial example exists \citep{wilkinson2004parable}. The outcome of this process is a complex ecosystem where species are `fit' for their environments. The origin of this fit is not cumulative selection nor evolution. Instead it is similar to a sieve, because there is no generation of variation (condition 1). 

Perhaps a more realistic scenario, which I will call \emph{ecological fitting}, is the migration of a new species to an established ecosystem. The entity is the ecosystem, its features are its species, their populations, the abiotic resources and the inter-species and species-environment interactions. Such a list would be extremely difficult to determine experimentally but is very accessible in models like multi-species Lotka-Volterra \citep{arthur2017entropic}. The continuous arrival of migrants is the source of variation (condition 1). 

The maintenance of the ecosystem state at or near present conditions is called ecosystem \emph{stability} while ecological \emph{resilience} is the ability of an ecosystem to resist or recover from disturbances and return to a state close to the previous one. For the purpose of assessing if cumulative selection is operating, I identify a combination of stability and resilience with persistence. The directional influence of ecosystem features, especially diversity, on stability and resilience has long been debated \citep{holling1973resilience}. For our purposes it is enough to know that they have an effect to satisfy condition 2, and that the introduction of a very small number of migrants will not dramatically change ecosystem stability and resilience, condition 3 \citep{hooper2005effects}. 

Thus, ecological fitting through the continuous arrival of migrants can explain ecosystem adaptation for persistence. Again, this does not mean co-evolution or niche construction do not happen, rather it means ecological fitting via migration is a plausible adaptive explanation for observations.

I note that parasites, like the Dutch elm fungus, can cause huge changes in ecosystems in a relatively short time. I have mostly been discussing small changes and this kind of long jump \citep{kauffman1987towards} apparently causes a violation of CS Condition 3. However, just as the `tempo' of evolution \citep{gould1977punctuated} is not restricted in most contemporary accounts of ENS, so the tempo of cumulative selection is not necessarily slow nor constant. As in punctuated equilibrium there will be continuity and correlation over short times and the definition of a `long' jumps depends on the timescale under consideration.

Finally, consider the global ecosystem, here equated with Gaia. This is claimed to undergo Sequential Selection by \cite{lenton2018selection}. This mechanism posits that global biota experience periodic collapses, after which a new global biota arises. By definition, the global ecosystems that produce an extensive fossil record, or the one we are likely to live in, are the persistent ones. This version of Sequential Selection, taken from a strict reading of \citep{lenton2018selection}, is not cumulative selection. There is no continuity between consecutive states, violating condition 3 and preventing the process from being \emph{cumulative}. 

As emphasised by \citep{arthur2022selection} (and probably intended by \cite{lenton2018selection}), Sequential Selection \emph{with Memory} suggests if there is continuity between successive states e.g. through continuity of biogeochemical cycles \citep{doolittle2018processes} or maintenance of biodiversity in refugia \citep{arthur2024does}, then the process can be cumulative. In this case we have: condition 1, variation by periodic mass extinction, condition 2, differential persistence by global ecosystem stability and resilience and condition 3, memory via continuity of species and evolutionary adaptations through extinction events. 

This is quite far from standard ENS. What is the analogue of populations and reproduction? Other authors have gone to great lengths to find some mapping of this situation back to standard ENS e.g. \citep{doolittle2024darwinizing} with arguable success \citep{arthur2025does}. The advantage of cumulative selection is that we can posit an adaptationist explanation for the observation of Gaia, that is, the `fit' of extant life and the environment at the global scale, without such a mapping. It might ultimately be an incorrect explanation, but using this framework we can make the claim \citep{lenton2018selection}, study the micro level mechanisms driving it \citep{arthur2022selection}, understand its consequences \citep{arthur2024does} and make predictions \citep{arthur2023gaian}.

\subsubsection{Neural Networks:} Some machine learning algorithms fit into this framework. In broad strokes, a machine learning task, like classifying images of dogs and cats, is translated into a `loss function' $L(y, \hat{y})$ where $y$ is the required output and $\hat{y}$ the prediction \citep{hastie2009elements}. The loss function is minimised by predictions, $\hat{y}$, close to the truth $y$. A biologist might prefer to think of a fitness function $F = -L$ which is \emph{maximised}, the two are equivalent. A prediction algorithm is just a very flexible function $f(x; s) = \hat{y}$ which transforms inputs $x$ into predictions $\hat{y}$ based on its state $s$. The state is typically specified by a very large set of parameters. In a neural network it is the weights and biases connecting the layers and controlling the activations of the neurons.

One way of performing this optimisation is using \emph{stochastic gradient descent} (SDG). The basic equation is
$$
s \rightarrow s - \eta \nabla L
$$
which means that the state $s$ is updated by a small amount, $\eta$, in the direction of steepest descent, $\nabla L$, over a large number of iterations. When the current state yields predictions which are far from optimal, the gradient is large and the changes are large. When the state is close to optimal the gradient is small and the steps are small. The method is called \emph{stochastic} because at each iteration only a random selection of training data is used to compute the gradient of the loss function.

The entity is the neural network, the features are the parameters and the state is the current set of parameter values
\begin{enumerate}
    \item The update equation $s \rightarrow s - \eta \nabla L$ induces random variation in the parameters
    \item States far from optimal (large losses) produce larger gradients. States close to optimal produce smaller ones. Exactly optimal features produce zero gradients (though this is unlikely in practice). Thus a parameter range $s_i \pm \delta s_i$ closer to optimal is more \emph{persistent} than one far from optimal.
    \item The rate of variation is controlled by $\eta$, called the learning rate, which must be small and in fact, is often made to decrease as the optimum is approached.
\end{enumerate}
So the SDG algorithm is performing cumulative selection on the parameters.

One objection to identifying SDG as a form of cumulative selection is that the variation is not completely random, it is biased by the gradient of the objective function. Even though SDG does not function via purely random variation, allowing forms of \emph{biased} search to be encompassed by cumulative selection seems to do little harm. A biological case would be selective breeding or domestication, where variation in the offspring is controlled by artificial selection. Such cases still satisfy Lewontin's conditions, though we would not call them \emph{natural} selection.

This application is included mostly to stimulate ideas. Optimisation theory often takes inspiration from biology, e.g. genetic algorithms \citep{holland1992genetic}. Understanding new forms of biological change could lead to cross-fertilisation between disciplines, new algorithms or perhaps even motivation to look for novel selection processes of biological interest.

\begin{landscape}
\begin{table}
\centering
\begin{tabular}{|p{3.25cm}||p{3.25cm}|p{3.25cm}|p{3.25cm}|p{3.25cm}|p{3.25cm}|}
\hline
\textbf{Process} & \textbf{System} & \textbf{Features} & \textbf{Generator of variation} & \textbf{Cause of differential persistence} & \textbf{Mechanism of correlation} \\
\hline\hline
Individual lifecycle & Individual & Genes & \textbf{None} & Environment & Survival \\
\hline
Genetic drift & Population & Genes & Neutral mutations & \textbf{None} & Reproduction \\
 \hline
Random Search & Lottery ticket & Numbers & Random guesses & Match to target & \textbf{None} \\
\hline
Sieve & Population & Sizes & \textbf{None} & Mesh size & Survival \\
\hline
Selection by Survival & Population & Species & \textbf{None} & Extinction & Survival \\
\hline
Ecological colonisation & Ecosystem & Biota, Abiota, interactions & \textbf{None} & Resource competition & Reproduction, recycling, resource flow \\ \hline
Sequential Selection & Gaia & Biota, Abiota, interactions & Reset + Ecological colonisation & Resilience to perturbation & \textbf{None}\\ \hline
\end{tabular}
\caption{Examples of entities subject to changes which \emph{are not} cumulative selection.}\label{tab:neg}
\end{table}
\end{landscape}

\begin{landscape}
\begin{table}
\centering
\begin{tabular}{|p{3.25cm}||p{3.25cm}|p{3.25cm}|p{3.25cm}|p{3.25cm}|p{3.25cm}|}
\hline
\textbf{Process} & \textbf{System} & \textbf{Features} & \textbf{Generator of variation} & \textbf{Cause of differential persistence} & \textbf{Mechanism of correlation} \\
\hline\hline
ENS & Population & Phenotypes & Mutation, recombination & Environment & Reproduction \\
\hline
Cumulative Search & Lottery ticket & Numbers & Random guesses on unmatched sites & Match to target & Fixation of correct numbers \\
\hline
Sequential Selection with Variation & Clade & Diversity & Speciation & Environment & Survival \\
\hline
Persistence Through Time & Pando & Ramet locations & Propagation & Environment & Proximity \\
\hline
Holobiont Selection & Population & Microbiome size & Mixing & Microbiome-host interaction & Vertical/Horizontal transmission \\
\hline
Rein Control & Gaia & Albedo & Daisy coverage & Temperature effect on growth & Reproduction \\
\hline
Ecological fitting & Ecosystem & Biota, Abiota, interactions & Migration & Resilience to perturbations & Reproduction, recycling, resource flow \\ \hline
Sequential Selection with Memory & Gaia & Biota, Abiota, interactions & Reset + Ecological colonisation & Resilience to perturbation & Refugia, low level species persistence, seed banks\\ \hline
\end{tabular}
\caption{Examples of entities subject to changes which \emph{are} cumulative selection.}\label{tab:pos}
\end{table}
\end{landscape}

\section{Discussion}\label{sec:disc}

This paper describes cumulative selection, a type of stochastic search that can generate complex adaptations in sub-exponential time by accumulation of persistence enhancing variation. I have given a general set of conditions for some entity to be subject to cumulative selection and a number of examples of systems which do and do not satisfy these conditions. 

This work draws on many ideas that have been advanced in the literature: Persistence Through Time \citep{bouchard2004evolution}, Selection by Survival \citep{ford2014natural}, Stability based Sorting \citep{toman2017stability}, ITSNTS \citep{doolittle2018processes}, Sequential Selection \citep{lenton2018selection}, as well as older ideas around adaptive walks \citep{metz1995adaptive, kauffman1987towards}. The more recent approaches have tended to either modify Lewontin's recipe \citep{bouchard2004evolution, bourrat2014survivors, charbonneau2014populations, papale2021evolution} or describe alternative  processes \citep{ford2014natural, lenton2018selection}. The key contribution differentiating this work is that I have attempted a \textbf{generalisation} of ENS not an alternative to or extension of it. ENS is a particular example of cumulative selection, but there are others.

 There are two main advantages to this point of view. The first is that \textbf{cumulative selection can provide an adaptive account of observed complexity}. If we identify a system that satisfies the conditions for CS we do not get all the machinery of paradigmatic-ENS but we do still retain \emph{some} of its  explanatory power, in particular the ability to account for the generation of complex or fine-tuned features in short times. This means we can posit CS as a plausible mechanism to explain some observation of complexity or fine-tuning in the system of interest. 
 
This does not mean that such explanations are the only correct ones. For example, the theory of constructive neutral evolution (CNE) \citep{stoltzfus1999possibility, lukevs2011neutral, brunet2018generality, catherall2025constructive} provides a convincing argument that neutral change, rather than adaptation, is responsible for the complexity of cellular machinery and potentially other biological and non-biological systems. The correctness of a CS account of complexity versus a neutral one, or any other, is an empirical question. Nor does this mean mean that CS automatically determines the appropriate level of analysis, definition of persistence and so forth. These must be specified by the investigator and checked against the conditions.

The second benefit of this perspective is that \textbf{cumulative selection allows us to study biological systems in terms of our desired `focal unit'}. This lets us largely sidestep debates \citep{papale2021evolution, doolittle2024darwinizing, zhang2025defending} over the definition of reproduction and populations. We can let ENS, or paradigmatic-ENS, be something recognisable to Darwin, while recognising that other processes, like Selection by Survival or Sequential Selection, can also explain complexity. Pando or Gaia can be doing cumulative selection but not ENS. 

While some authors believe being able to call processes of change `Darwinian' is rhetorically vital \citep{doolittle2024darwinizing}, letting go of this requirement is a conceptual simplification. Reproduction can be reproduction, growth can be growth, one does not have to encompass the other (as in \citep{godfrey2009darwinian}). Populations can be used when relevant (ENS) and not when they are awkward (a lottery ticket as a correlated population of numbers, an ecosystem as a population of interactions etc.) Identifying a system and studying how it varies allows us to determine if some feature of it can be plausibly explained by differential persistence. Deeper study can then identify the `micro' causes of variation and differential persistence.

An explanation for the sex-ratio in terms of both ENS and CS on populations was given in Section \ref{sec:examples}. Here the higher-level CS description is largely redundant and doesn't offer any new insight. Population sex ratio is simply an aggregate of individual features. We know the micro details and can apply the more powerful (but more restrictive) theory of ENS. However there can be emergent concepts of fitness or persistence that can apply which are not reducible to a summation of individual values (similar to \cite{okasha2006evolution}'s $\text{fitness}_1$ and $\text{fitness}_2$). In the examples given in Section \ref{sec:examples} the clearest case of this is ecosystem evolution. Ecosystem resilience and stability are clearly properties of the whole not reducible to aggregate properties of individual organisms or species. Nonetheless cumulative selection still provides a theoretical basis for making an adaptationist argument to explain their features.

New facts force us to update and expand existing theories. To take one example, corals are modular \citep{tuomi1989hierarchical}, clonal \citep{clarke2012plant}, sexually reproducing \citep{harrison2010sexual} holobionts \citep{roughgarden2020holobiont} and niche constructors \citep{kylafis2011niche}. Forcing all of these possibilities for adaptation to conform to classical ENS formulations is unnecessarily restrictive. Cumulative selection on the other hand is a general description of adaptation by random variation which encompasses not only evolution by natural selection but other interesting biological and even technological processes. This moves focus away from debates about the proper interpretation of ENS and towards more important issues - the origins of adaptations.

\section*{Acknowledgements} 
Thanks to Arwen Nicholson, David Wilkinson, Tyler Brunet and two anonymous referees for their helpful comments on the manuscript.

\appendix
\section{Model Solutions}\label{sec:app}

We consider sequences of $L$ characters from an alphabet of size $b$. There is some target sequence $s_T$ and $N$ sequences $s_i$. Each of the $N$ sequences is subject to some kind of random variation every time step and we are concerned with how long it takes one of these sequences to reach the target.

\subsection{Random Sequences}

Let us first consider the non-cumulative version of the model. At each step we randomly generate $N$ new sequences $s_i$. The previously discovered letters are lost, and the only way to discover the target is all at once.
A random sequence has a $\frac{1}{b^L}$ chance of matching the target, so $N$ independent sequences have probability of matching the target equal to 
\begin{equation}
    p(N) = \frac{ N }{b^L}
\end{equation} 
Using a standard trick we either match on the first trial or fail the first trial and match in the expected number of trials, since this is a memoryless process. This gives
\begin{align}
T &= p + (1-p)(T+1) \\ \nonumber
T &= 1/p = \frac{b^L}{N}
\end{align}
trials on average. For $N=1$ we have $T = b^L$.

\subsection{Recursions}

Now consider a cumulative search with a population $N=1$. At each step we reproduce each character from the previous step with a probability $\mu$ to change a character that doesn't match with the target and a probability $\nu$ to change a matching character. For simplicity we set $\nu = 0$ at first, so that once a character is matched it stays fixed.

Call the expected number of generations needed to match $n$ characters in the target $T(n)$. $T(0) = 0$ by definition. $P(n,k)$ is the probability of making $k$ matches in $n$ positions. The expected number of trials required to match $n>1$ characters is at least 1, since we have to do at least one new trial; plus the chance of matching no characters multiplied by the wait time to match $n$, $P(n,0)T(n)$; plus the chance of matching one character times the remaining wait time to match $n-1$, $P(n,1)T(n-1)$ and so on up to $P(n,n) T(0) = 0$ the, very unlikely, case where we match all the characters in the next turn.
\begin{equation}
    T(n) = 1 + \sum_{k=0}^n P(n,k) T(n-k)
\end{equation}

For the unordered case
\begin{equation}
P(n,k) = \binom{n}{k} p^k q^{n-k}
\end{equation}
where $p$ is the chance of success at a single site and $q = 1-p$. When mutation happens at a site with probability $\mu$ and there is a probability $1/b$ to choose the right character we have 
\begin{equation}
    p = \frac{\mu}{b}
\end{equation}
For the ordered case we have the same $p$ and
\begin{equation}
P(n,k) = p^k q
\end{equation}
meaning we find exactly $k$ matches in order, then fail at the $(k+1)^{th}$.

\subsection{Ordered Solution}

When the matching characters must be discovered in order, from left to right say, we can solve the recurrence exactly. First move all the $T(n)$ terms to the left
\begin{align}
    T(n) &= 1 + \sum_{k=0}^n p^kq T(n-k) \\
    &= 1 + T(n)q + q\sum_{k=1}^n p^k T(n-k)\\
    (1-q) T(n) &= 1 + q\sum_{k=1}^n p^k T(n-k)
\end{align}
Define the difference operator (similar to a discrete derivative)
\begin{equation}
D(n) = T(n) - T(n-1)
\end{equation}
and compute $(1-q)(T(n) - T(n-1))$ to give
\begin{align}
    (1-q) D(n) &= q\sum_{k=1}^{n-2} p^k D(n-k) + qp^{n-2} 
\end{align}
using $T(1) = 1/p$. We attempt a solution where $D(n) = D$ for $n\geq2$ 
\begin{align}
    D(1-q) &=  Dq \sum_{k=1}^{n-2} p^k  + qp^{n-2} \\
\end{align}
Using the identity
\begin{align}
    \sum_{k=1}^{n-2} p^k = \frac{p(1-p^{n-2})}{1-p}
\end{align}
gives
\begin{align}
    D(1-q) =  Dq \frac{p(1-p^{n-2})}{1-p}  + qp^{n-2} 
\end{align}
which simplifies to   
\begin{align}
D = \frac{q}{p}
\end{align}

Using this, $D(1)=1/p$ and $T(L) = \sum_{n=1}^L D(n)$ gives
\begin{align}
    T(L) &= \sum_{n=2}^L D(n) + D(1)\\
    &= (L-1)\frac{q}{p} + \frac{1}{p} \\
    &= \frac{ (L-1) - p(L-1) + 1}{p} \\
    &= \frac{L}{p} - (L-1)
\end{align}
which is the naive expectation where we simply wait for every term independently, giving $\frac{L}{p}$, with a small correction to account for when we happen to get more than one character correct at a time.

\subsection{Unordered Solution}

When the correct characters can be found in any order, the recurrence is quite difficult to solve. It can be rearranged into different forms, but what we really want is the dependence of $T$ on $L$. We can approximate it by considering the state when there are $d$ characters remaining. The probability of finding the target in the next generation is $p^{L-d}$, if we don't allow back mutation. Assuming $\mu$ is small, the probability of getting multiple correct characters in a single generation can be neglected. The probability of any of the $d$ remaining positions matching in the next generation is then $dp$.

This means we wait $\tau(d) = \frac{1}{dp}$ generations to make one match when $d$ matches are needed. To match all the remaining entries, starting from $L$ missing, we therefore have to wait
\begin{align}
    T(L) = \sum_{d=0}^L \tau(d) = \frac{1}{p} \sum_{d=0}^L \frac{1}{d}
\end{align}
We have
\begin{equation}
\sum_{d=0}^L \frac{1}{d} = H_L
\end{equation}
Where $H_L$ is the $L^{th}$ harmonic number. A standard and very accurate approximation even for small $L$ is $H_L \simeq \ln L + \gamma$ where $\gamma = 0.5772156649...$ is the Euler-Mascheroni constant. Therefore
\begin{equation}
T(L) \simeq \frac{ \left( \ln L + \gamma \right) }{p}
\end{equation}
The key thing is that the time is logarithmic in $L$. This is \emph{much} faster than the ordered case which is linear in $L$, for example if $L=10^9$ then $\ln (10^9) = 9 \ln 10 \simeq 21$.

\subsection{$N>1$}

\begin{figure}
    \centering
    \includegraphics[width=0.85\linewidth]{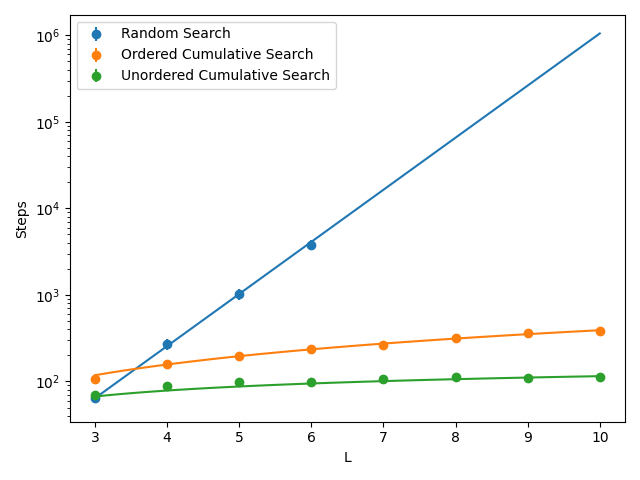}
    \caption{Simulations of the cumulative search process for $N=1$ compared to the analytical results above. Using $\nu=0, \mu=0.1, b=4$. Each point averages over 200 searches.}
    \label{fig:fig1}
\end{figure}
\begin{figure}
    \centering
    \includegraphics[width=0.85\linewidth]{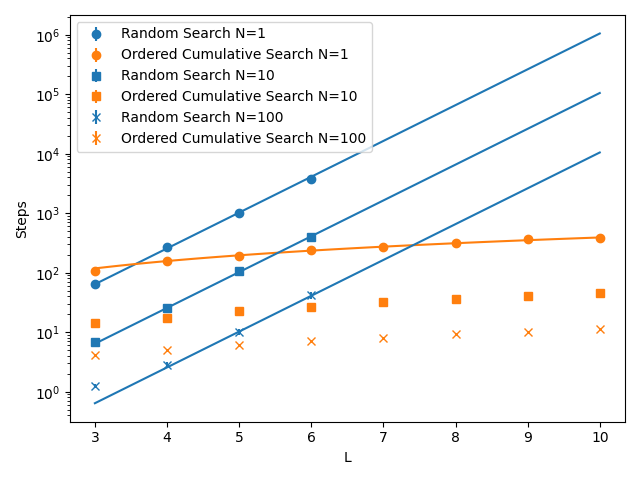}
    \caption{Simulations of the cumulative search process for $N>1$ using $\nu=0, \mu=0.1, b=4$. Each point averages 200 searches.}
    \label{fig:fig2}
\end{figure}

The solutions for $N>1$ involve choosing the \emph{maximum} matching sequence among $N$ instances which is somewhat more complicated to analyse and does not really add a huge amount to our story. Figure \ref{fig:fig1} shows some simulations of the process for $N=1$ confirming our analysis, in particular the accuracy of the approximation made for the unordered case. Figure \ref{fig:fig2} shows some examples of $N>1$ making the basic point from the text that for small $L$ and large $N$ the random search can be faster than the cumulative one, but at larger $L$ the order always switches for any $N$.

\subsection{Error Threshold}

We now consider what happens in the cumulative model if, when a match is found, there is a small probability $\nu$ to lose it. The probability to keep a correct match is the probability to not mutate or the probability to mutate, but for that mutation to reproduce the correct character,
\begin{equation}
r = (1-\nu) + \frac{\nu}{b}
\end{equation}
The goal is to find the value of $\nu = \nu_c$ where the rate of this `back mutation' is high enough that cumulative selection is impossible.

\subsubsection{Unordered}
For the unordered case we can treat every entry independently. Let $\pi$ be the probability the site is correct in the long term. At this stationary point we have a balance between the probability of transitioning away from the correct value and the probability of transitioning to it. Recall $\nu$ is the the probability to mutate a fixed entry and $\mu$ is the probability to mutate an unfixed one. The probability to go from an incorrect state to a correct one is
\begin{equation}
(1-\pi) \frac{\mu}{b} 
\end{equation}
The probability to go from a correct to an incorrect is
\begin{equation}
\pi (1-r)
\end{equation}
These balance at stationarity
\begin{align}
\pi (1-r) &= (1-\pi) \frac{\mu}{b} \\
\pi &= \frac{\frac{\mu}{b}}{ \frac{\nu(b-1)}{b} + \frac{\mu}{b}}
\end{align}
The probability for $L$ sites to be correct at stationarity is $\pi^L$ with waiting time $\pi^{-L}$. The exponent can be offset only if $\pi$ is sufficiently large. Using the expansion $\exp(-x) \simeq (1 - x)$ for small $x$ note that if $\pi \sim 1 - (1/L)$ then $\pi^{-L} \sim \frac{1}{e}$ which is a `reasonable' wait time. Thus to find the sequence in reasonable time requires
\begin{align}
\pi \geq 1 - (1/L)
\end{align}
or solving for the critical value of $\nu_c$
\begin{align}
    \nu_c =  \frac{\mu}{b-1}  \frac{1}{L-1} 
\end{align}
The rate of back mutation can't be higher than the rate of forward mutation divided by the the alphabet size and target length.

\subsubsection{Ordered}
Let the random variable $X$ be the length of the prefix in the next generation. The tail-sum formula for the expectation of $X$ is
 \begin{equation}
E[X] = \sum_{i \geq 1} P(X \geq i)
 \end{equation}
 If the current prefix is of length $m$, then for $i \leq m$, $P(X \geq i) = r^i$ and for $i>m$, $P(X \geq i) = r^m p^{m-i}$. 
  $$
E[X] = \sum_{i \geq 1}^m r^i + r^m \sum_{i = 1}^{L-m} p^i
 $$
We need the prefix to eventually be of length $L$, so the second term vanishes and in the limit of large $L$ the first term is approximately $\frac{r}{1-r}$. If expected length is less than $L$ then a prefix this long cannot be sustained due to the rate of back mutation in which implies
$$
\frac{r}{1-r} < L
$$
Solving for $\nu_c$ gives
\begin{align*}
    \nu_c &= \frac{1}{L+1}\left(\frac{b}{b-1} \right)
\end{align*}
This just depends on $L$ with a very weak dependence on $b$ and no dependence on $\mu$.

\subsection{Complicating the model}

There are various ways we could make this model more realistic \\

\noindent \textbf{Fixation Time:} We are interested in the search time, but we could also consider the time spent copying numbers onto $N$ new lottery tickets, or, in the genetic case, for the mutation to become ubiquitous in a population. This depends on the selective advantage of the mutation. In standard population genetic models the fixation time scales like $\log N$ for a beneficial mutation or like $N$ for a neutral mutation \citep{ewens2004mathematical}. Taking the worst case, we could add a factor proportional to $N$ to the total time for every generation. This creates a tension between search time (faster for large $N$) and fixation time (slower for large $N$) where some optimum $N$ could be found. The back mutation rate $\nu$ could be framed in a similar way, as a combination of a baseline mutation probability and purifying selection against back mutation. This kind of extension is beyond the scope of what we want to do since we are mostly interested in the case $N=1$.\\

\noindent \textbf{Continuous variables:} Using a fixed alphabet is natural when considering a sequence of genetic bases or amino acids and makes the analysis much simpler. However we will also consider states which are specified by continuous parameters like size or latitude. We can make continuous variables discrete by binning them, measuring a parameter $s$ in increments of $\delta s$. A much more mathematically interesting approach would be to model this situation as a diffusion process \citep{ewens2004mathematical}, but this is beyond the scope of this paper. \\

\noindent \textbf{Variable sized alphabets and sequences:} On the other extreme, for something like the nitrogen cycle, the letters of the sequence could be binary variables indicating the presence/absence of a chemical species. We could also allow each letter to be chosen from a different alphabet e.g. consonants in the first position, vowels in the second and so forth. We could allow the sequences to get longer and shorter by inserting or deleting letters with some probability.\\

\noindent \textbf{Complex targets:} In case the reader is concerned that the sequence model is aiming at a predetermined target, this is simply to make the model easy to understand. If any sequence of 23 letters that made a sensible English phrase were permitted as a target this would make the `persistence landscape' more interesting, but much more difficult to analyse. As already mentioned we are solving the inverse problem: is it plausible that the observed target could be found by a random process? Alternatively, in the forward direction, we could consider the target sequence as representing a single peaked `fitness landscape', which is an approximation to a real system over a relatively short time-scale or describes selective breeding \citep{arthur2017decision}.\\ 

\noindent Rather than pursue reality, we consider here a tractable toy model, meant to display cumulative selection in the simplest possible setting and don't consider these and other possible extensions.

\subsection{List of Symbols}
\begin{enumerate}
    \item $L$ sequence length
    \item $b$ alphabet size
    \item $N$ number of sequences
    \item $s_i$ a sequence of $L$ letters from the alphabet, $i \in [1,N]$
    \item $s_T$ the target sequence / wining ticket / lock combination
    \item $\mu$ mutation probability of matched characters
    \item $\nu$ mutation probability of unmatched characters
    \item $p$ probability for an matching character to mutate to a non-matching one
    \item $r$ probability for a matching character to stay matching
    \item $T(n)$ the expected number of generations needed to match $n$ characters
    \item $\pi$ stationary probability of a match in the unordered case
\end{enumerate}

\bibliographystyle{plainnat}
\bibliography{references}

\end{document}